\documentclass[twocolumn,aps,showpacs,amsmath,amssymb]{revtex4-1}
\usepackage{graphicx}
\usepackage{dcolumn}
\usepackage{bm}
\usepackage{psfrag}
\usepackage{datetime}
\usepackage{color}
\usepackage{hyperref}
\def\beq{\begin{equation}}
\def\eqn{\end{equation}\noindent}
\def\eqnr{\end{eqnarray}\noindent}
\def\beqr{\begin{eqnarray}}
\def\epsilon{\varepsilon}
\begin{document}
\title{Revisiting linear augmentation for stabilizing stationary solutions: potential pitfalls and their application}
\author{Rajat Karnatak}
\email{rajat@pks.mpg.de}
\affiliation{Nonlinear Dynamics and Time Series Analysis Research Group, Max--Planck--Institute for the Physics of Complex Systems,
N\"othnitzer Str. 38,
01187 Dresden,
Germany}
\begin{abstract}  
Linear augmentation has recently been shown to be effective in targeting desired stationary solutions, suppressing 
bistablity, in regulating the dynamics of drive response systems and in controlling the dynamics of hidden attractors. The simplicity 
of the procedure is the highlight of this scheme but at the same time questions related to its general applicability 
still need to be addressed. Focusing on the issue of targeting stationary solutions, this work demonstrates instances 
where the scheme fails to stabilize the required solutions and leads to other complicated dynamical scenarios. Appropriate 
examples from conservative as well as dissipative systems are presented in this regard and potential applications for relevant 
observations in dissipative predator--prey systems are also discussed.
\end{abstract}
\pacs{05.45.Ac, 05.45.Pq, 05.45.Xt}
\maketitle
\section{Introduction}
Studies on coupled nonlinear systems have explored a wide variety of emergent dynamical phenomena, namely synchronization~\cite{Pikovsky2001}, 
oscillator suppression~\cite{Saxena2012,Koseska2013}, multistability~\cite{Feudel2008}, hysteresis~\cite{Prasad2005}, extreme--events~\cite{Ansmann2013,Karnatak2014b} etc. 
which can be exploited in applications, to model natural phenomena or in regulating the system behavior for instance. Controlling dynamical systems towards a desired behavior is an 
important research topic in nonlinear sciences~\cite{Scholl2008}. Starting with chaos control~\cite{OGY1990,Pecora1990,Pyragas1992,Peng1996}, 
research in this domain now also extends towards control of multistability~\cite{Pisarchik2014}, patterns and spatio--temporal chaos~\cite{Gang1994,Boccaletti1999}, 
noisy systems~\cite{Balanov2004,Hauschildt2006}, methods of stabilizing unstable stationary solutions~\cite{Bar-Eli1984,Saxena2012,Koseska2013} etc. 
A greater understanding of these different regulatory aspects have greatly contributed towards development of related novel and highly efficient procedures. 
Considering noninvasive (without changing the intrinsic system parameters) mechanisms leading to stabilization 
of stationary solutions, oscillator suppression via coupling nonlinear systems has been discussed extensively in literature 
(see Refs.~\cite{Saxena2012,Koseska2013} for detailed reviews). This suppression is majorly observed as a consequence of parameter heterogeneity between 
the coupled units~\cite{Ermentrout1990,Aronson1990,Mirollo1990}, presence of time--delayed~\cite{Reddy1998,Atay2003}/conjugate 
variables~\cite{Karnatak2007} in the coupling function or through dynamic coupling~\cite{Konishi2003}.

Recently, linear augmentation has also been suggested as another practical alternative leading to oscillator suppression, 
achieved by coupling systems to a linear feedback consisting of a simple decaying function~\cite{Pooja2011}. Lately, studies have also used linear augmentation 
effectively for controlling bistability~\cite{Pooja2013}, dynamics of a drive response system~\cite{Pooja2014} and in controlling the dynamics of 
hidden attractors~\cite{Sharma2015}. With respect to stabilizing stationary solutions, Ref.~\cite{Pooja2011} discussed results for an augmented Lorenz system where 
either the stationary solutions of the original system or those of the augmented system could be 
stabilized by picking an appropriate feedback function; former being quite relevant from an application perspective. The paper also presented some 
parameter space scans highlighting the regimes where linear augmentation works and where it does not, which although is instructive but is also very system 
specific at the same time. At this point, one must question the ability of linear augmentation towards stabilizing the stationary solutions in a more 
general sense, namely the systems, parameter settings and coupling configurations where the scheme works and where it does not? 
In this paper, we will look at some simple 
examples of linearly augmented systems demonstrating the fact that we need to be quite careful before picking linear augmentation in applications. 
These examples illustrate that there could be situations where even picking an appropriate feedback function does not guarantee that the required stationary 
solutions will be necessarily stabilized. We will see that the mechanism appears to be highly dependent on the intrinsic properties of the oscillators in consideration, 
the stationary solutions we want to target, and also on how these systems are augmented by/coupled to the feedback. Furthermore, we will also  discuss instances where the 
failures of the procedure can be exploited in applications. 

The manuscript is arranged as follows: Linear augmentation is introduced in the 
following Sec.~\ref{sec:linear_augmentation}. In Sec.~\ref{sec:examples}, we will have a look at results for linearly augmented conservative and dissipative 
dynamical systems. In Sec.~\ref{sec:harmonic}, results for partially and fully augmented harmonic oscillator are presented, Sec.~\ref{sec:duffing} discusses results 
for conservative Duffing oscillator under similar augmentations, and in Sec~\ref{sec:populations}, we will have a look at the behavior of augmented dissipative 
population models and also briefly discuss possible applications for certain observations in these systems. The manuscript concludes with a summary of results in 
Sec.~\ref{sec:summary}. Additional details on certain dynamical aspects of harmonic oscillator and Duffing system which were excluded from main manuscript 
are provided in Appendices~\ref{sec:harmonicappendix}, \ref{sec:duffinghysteresisappendix}, 
and \ref{sec:duffingappendix}.
\section{Linear augmentation}\label{sec:linear_augmentation}
General representation of a linearly augmented dynamical system is,
\begin{equation}
 \left.\begin{array}{l}
\dot{{\bf x}} = {\bf f(x)}+ \boldsymbol{\varepsilon}u\\
\dot{u} = -k u -\boldsymbol{\varepsilon}(\mathbf{x-b})^T.\\
\end{array}\right\}
\label{eq:i15}
\end{equation}\noindent
where the column vector $\mathbf{x}=[x_1, x_2, \ldots, x_N]^T \in \mathbb{R}^N$ ($[...]^T$ corresponds to the transpose) 
contains the systems variables, and $u$ is the augmentation variable. 
$\boldsymbol{\varepsilon}=[\varepsilon_1, \varepsilon_2, \ldots, \varepsilon_N]^T \in \mathbb{R}^N$ is a column vector with information 
regarding the coupling strength of the interaction between the dynamical variables and $u$; augmentation/coupling term corresponding to the $i^{th}$ 
component $x_i$ is 
$\varepsilon_i u$ $\forall$ $i=1,2, \ldots, N$, and $\varepsilon_i = 0$ if $x_i$ is not coupled to $u$. $\mathbf{b} \in \mathbb{R}^N$ is an 
arbitrary vector and $k$ is the decay constant~\cite{Resmi2010} 
which could be used to control the transient time leading to stabilization of stationary solutions~\cite{Pooja2011}.
Vector $\mathbf{b} = \mathbf{x^*}=[{x_1}^*, {x_2}^*, \ldots, {x_N}^*]^T \in \mathbb{R}^N$ where $\mathbf{x^*}$ satisfies 
$\dot{{\bf x}}|_{\mathbf{x=x^*}} = {\bf f(x^*)}= \mathbf{0}$ if we want to stabilize a stationary solution $\mathbf{x^*}$ of the original system. 
Substituting a value of $\mathbf{b} \neq \mathbf{x^*}$ can stabilize stationary solutions of 
augmented system for which $\mathbf{\dot{X}}=[\dot{x_1}, \dot{x_2}, \ldots, \dot{x_N}, \dot{u}]^T \in \mathbb{R}^{N+1} = \mathbf{0}$. 
The term $\boldsymbol{\varepsilon}(\mathbf{x-b})^T$ gives the dot product of the corresponding column vectors.

In the following, we will look at examples of augmented conservative as well as dissipative dynamical systems which highlight the 
limitations of this procedure. The terms \textsl{augmentation/augmented} and \textsl{coupling/coupled} have been 
used synonymously in the following text.
\section{Examples}\label{sec:examples}
Here we will discuss the instances of systems controlled via linear augmentation. We will first look 
at two examples of conservative systems, namely the harmonic oscillator and conservative 
Duffing oscillator where linear augmentation will be used to stabilize their stationary solutions.
\subsection{Harmonic oscillator}\label{sec:harmonic}
\begin{figure}
\scalebox{0.5}{\includegraphics{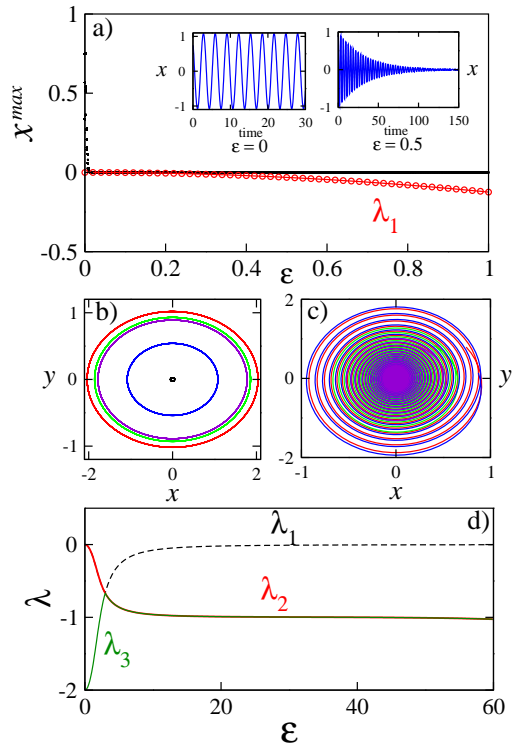}}
\caption{(Color online) a) Bifurcation diagram (black dots), largest 
eigenvalues (red symbols) and the time series of $x$ variable before and after the stabilization of origin as inset. 
Corresponding phase space plots are in b) (no augmentation: $\varepsilon=0$) and c) (with augmentation: 
$\varepsilon=0.5$). d) Variation in the eigenvalues for higher values of augmentation strength $\varepsilon$. 
Parameter values for  calculation are $\omega=2$ and $k=2$.}
\label{fig:figharmonic}
\end{figure}
Equations describing a linearly augmented harmonic oscillator are:
\beqr
\nonumber
\dot{x} &=& y+\varepsilon u,\\\nonumber
\dot{y} &=& -{\omega}^2 x,\\
\dot{u} &=& -k u - \varepsilon x,
\label{eq:harmonic1}
\eqnr
where $\omega$ is the frequency of the oscillator, $k$ is the augmentation parameter, and $\varepsilon$ is the coupling strength. The first two 
equations governing the evolution of $x$ and $y$ correspond to the original harmonic oscillator dynamics. In the absence of augmentation, harmonic oscillator 
conserves total energy, which stays constant on the ellipses shown in Fig.~\ref{fig:figharmonic}~b). Each of these ellipses correspond to the system evolution 
following different initial values of position and momentum, and hence, different conserved total energies. Harmonic oscillator has $x^*=0,y^*=0$ as the only 
stationary solution and note that the augmentation term only appears in the rate equation of the position variable $x$ at this point. In case of a successful 
stabilization, the required stationary solution of the full system should be $(x^*,y^*,u^*)=(0,0,0)$ (\textsl{origin}) where the system effectively 
decouples from augmentation.

The characteristic eigenvalue equation at the origin is,
\beqr
(\lambda +k)(\lambda^2+\omega^2)+\varepsilon^2\lambda=0.
\label{eq:char.harmonic.1}
\eqnr
For $\varepsilon=0$, the eigenvalues for the full system are $\lambda_{1,2}=\pm i\omega$ which correspond to the non-hyperbolic stationary solution at the origin, 
and $\lambda_3=-k$ corresponding to the decay of the augmentation variable $u$; which evolves as $u(t)\propto exp(-kt)$ in this case. The parameter values for the following 
calculations were fixed at $\omega=2$, and $k=2$. For the evolution of the augmented system ($\varepsilon > 0$), the bifurcation diagram~\cite{bif_diagram} of the system 
with increasing $\varepsilon$ values is shown in Fig.~\ref{fig:figharmonic}~a)~(black dots). It is seen that with an increasing $\varepsilon$, the system 
which was conservative for $\varepsilon=0$ becomes dissipative and gets into a stable origin regime even for quite small values of $\varepsilon$. 
Rewriting Eq.~\ref{eq:char.harmonic.1} as,
\beqr
\lambda^3 + k\lambda^2 + (\omega^2+\varepsilon^2)\lambda + k\omega^2 = 0,
\label{eq:char.harmonic.2}
\eqnr
and applying the Routh--Hurwitz criteria (RHC)~\cite{Routh_Hurwitz_criteria}, we can deduce 
that the roots of this equation are all either negative or have negative real parts (in case of complex roots) $\forall$ $\varepsilon > 0$. Largest eigenvalues 
of the Jacobian (red symbols) are also plotted along with the bifurcation 
diagram in Fig.~\ref{fig:figharmonic}~a) which demonstrate the transition from oscillatory to stationary state for $\varepsilon >0$. 
Considering the behavior of this system for large $\varepsilon$, we can see that the largest eigenvalue 
$\lambda =\lambda_1 \to 0$ from Eq.~\ref{eq:char.harmonic.2} in this limit. Since the discriminant \cite{discriminant} 
of the cubic characteristic Eq.~\ref{eq:char.harmonic.2} is negative $\forall$  $\varepsilon \geq 0$, this implies that 
the system has one real eigenvalue and a pair of complex conjugate eigenvalues in this range. The negative real part of 
these complex eigenvalues for large $\varepsilon$ can therefore be estimated by equating the sum of all eigenvalues 
to the trace of the Jacobian $\text{tr}(J)$, giving $\text{Re}(\lambda_{2,3})=-k/2$. This further implies that as $\varepsilon \to \infty$, 
convergence to the origin gets slower although origin is stable in the entire $\varepsilon>0$ range and any change in stability 
will only occur as $\varepsilon \to \infty$ when $\lambda_1 = 0$. 

Now let us consider a more general case of an augmented harmonic oscillator given by,
\beqr
\nonumber
\dot{x} &=& y+\varepsilon_1 u,\\
\dot{y} &=& -{\omega}^2 x+\varepsilon_2 u,\\\nonumber
\dot{u} &=& -k u - \varepsilon_1 x - \varepsilon_2 y,
\label{eq:harmonic2}
\eqnr
where the augmentation now appears in the rate equations of both position $x$ and momentum $y$ with coupling strengths $\varepsilon_1$, $\varepsilon_2$ respectively. 
The eigenvalue equation in this case is,
\beqr\nonumber
(\lambda+k)(\lambda^2+\omega^2)+\lambda({\varepsilon_1}^2+{\varepsilon_2}^2)+
\varepsilon_1\varepsilon_2(1-\omega^2)=0.\\
\label{eq:char.harmonic.3}
\eqnr 
Substituting $\varepsilon_2=0$ and $\varepsilon_1=\varepsilon$ in Eq.~\ref{eq:char.harmonic.3} yields 
the dynamics of Eq.~\ref{eq:harmonic1}. Similarly, for $\varepsilon_1=0$ and $\varepsilon_2=\varepsilon$, 
we obtain a case where the system is only  coupled in the $y$ variable for which the characteristic 
Eq.~\ref{eq:char.harmonic.3} is exactly identical to Eq.~\ref{eq:char.harmonic.2}, and 
therefore the stability characteristics of the origin are identical and independent 
of whether the system is augmented in $x$ or $y$. In the previous example, we saw a situation where linear 
augmentation successfully stabilized the origin for the entire range of $\epsilon>0$. Now considering 
$\varepsilon_1=\varepsilon_2=\varepsilon$ (the system is similarly augmented in both 
variables), in which case Eq.~\ref{eq:char.harmonic.3} gives,
\beqr
\lambda^3 + k\lambda^2 + (\omega^2+2\varepsilon^2)\lambda + k\omega^2 + \varepsilon^2(1-\omega^2) = 0.
\label{eq:char.harmonic.4}
\eqnr
Using the RHC, it can be seen that this equation will have all negative eigenvalues iff 
$k\omega^2 + \varepsilon^2(1-\omega^2)>0$ which gives a stability regime of 
$0< \varepsilon < \varepsilon^*$ $\forall$ $\omega>1$ where $\varepsilon^*=\omega\sqrt{\dfrac{k}{\omega^2-1}}$, 
and for higher values of $\varepsilon$, RHC suggests appearance of positive eigenvalue/eigenvalues.
For large $\varepsilon$, we can get an estimate of largest eigenvalue 
$\lambda_1 \to \dfrac{(\omega^2-1)}{2} >0$ $\forall$ $\omega>1$. Since the discriminant is 
negative $\forall$ $\varepsilon > 0$, the remaining complex conjugate eigenvalue pair have a negative real 
part given by $\text{Re}(\lambda_{2,3})=(\text{Tr}(J)-\lambda_1)/2=-\left(\dfrac{k}{2}+\dfrac{(\omega^2-1)}{4}\right)$. 
Therefore, unlike in the previous example, we can see that origin here is unstable for large $\varepsilon$. 
The expression for $\varepsilon^*$ also shows that a higher value of $k$ can extend the coupling range for a stable origin. 
This result is the first instance of unexpected behavior as we would normally expect a 
higher coupling value to keep the origin stable. Furthermore, in the $\varepsilon>\varepsilon^*$ 
regime it is numerically observed that the trajectories escape to infinity which is also quite unexpected.

One of the primary reasons behind considering augmented harmonic oscillator in this study is the fact that it is highly 
solvable and therefore can provide necessary insights into the physical mechanisms behind the desirable as well 
as undesirable behaviors. It turns out that in case of a successful stabilization, this system represents a forced harmonic 
oscillator where the steady state solution (which is completely determined by the forcing) decays to the origin along with 
an exponentially decaying force. Similarly, the case where the trajectories escape to infinity corresponds again to a forced 
system but this time being driven by an exponentially diverging force which is analogous to a situation where energy is being pumped 
into the system. Therefore the steady state solution in this case diverges along with the diverging force explaining 
the unexpected behavior of escaping trajectories observed for the fully augmented system. Details of the 
calculations leading to these deductions are available in Appendix~\ref{sec:harmonicappendix}.
 \subsection{Duffing oscillator}\label{sec:duffing}
 \begin{figure}
\scalebox{0.57}{\includegraphics{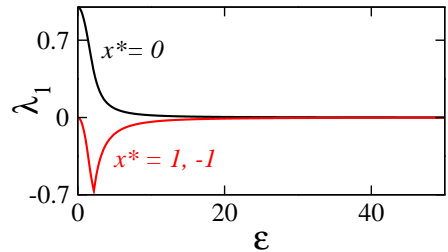}}
\caption{(Color online) Largest eigenvalue estimates for stationary solutions $(x^*,y^*)=(0,0)$ (black), 
$(x^*,y^*)=(\pm 1,0)$ (red) for partially augmented Duffing system.}
\label{fig:figduffeigenvals}
\end{figure}
General equations for a linearly augmented Duffing oscillator with no damping or forcing can be written as:
\beqr\nonumber
\dot{x}& = &y+\varepsilon_1 u,\\ \nonumber
\dot{y}& = &x-x^3 + \varepsilon_2 u,\\
\dot{u}& = &-k u - \varepsilon_1 (x-x^*) - \varepsilon_2 (y-y^*).
\label{eq:duffing1}
\eqnr
Uncoupled Duffing system has an invariant of motion (also the Hamiltonian of the system) given by $H(x,y)=y^2/2-x^2/2+x^4/4$ 
and stationary solutions: $(x^*,y^*)=(\pm 1,0),(0,0)$. The trajectories of this system evolve on the double well potential surface of $H(x,y)$ 
starting from different initial conditions for $\varepsilon_1=\varepsilon_2=0$. Similar to the previous example, for a successful stabilization, the 
required stationary solutions of the full system should be $(x^*,y^*,u^*)=(0,0,0)$ or $(\pm1,0,0)$ where the system effectively decouples from the 
augmentation. These solutions will be referred to as $(x^*,y^*)=(0,0)$ (origin) or $(\pm1,0)$ in the following.

For the system in Eq.~\ref{eq:duffing1}, the characteristic eigenvalue equation can be expressed as,
\beqr\nonumber
(\lambda+k)(\lambda^2+3{x^*}^2-1)+\lambda({\varepsilon_1}^2+{\varepsilon_2}^2)\\
+\varepsilon_1\varepsilon_2(2-3{x^*}^2)=0.
\label{eq:char.duffing.main}
\eqnr
Now similar to the harmonic oscillator example, considering partial augmentation 
with $\varepsilon_{1(2)}=\varepsilon$, and $\varepsilon_{2(1)}=0$ first, 
Eq.~\ref{eq:char.duffing.main} suggests that the stability characteristics 
for the stationary solutions are again independent of whether the system is being augmented 
in $x$ or $y$. For this partial augmentation, Eq.~\ref{eq:char.duffing.main} gives,
\beqr
(\lambda+k)(\lambda^2+3{x^*}^2-1)+\lambda{\varepsilon}^2=0.
\label{eq:char.duffing.main1}
\eqnr 
Substituting $x^*=0,\pm1$, and rearranging the terms, we can obtain the characteristic eigenvalue equations for these stationary 
solutions as,
\beqr
\lambda^3 + k\lambda^2 + (\varepsilon^2-1) \lambda -k=0,
\label{eq:char.duffing.main2}
\eqnr
for $(x^*,y^*)=(0,0)$ (hyperbolic for $\varepsilon=0$), and
\beqr
\lambda^3 + k\lambda^2 + (\varepsilon^2+2) \lambda +2k=0,
\label{eq:char.duffing.main3}
\eqnr
for $(x^*,y^*) = (\pm 1,0)$ (non hyperbolic for $\varepsilon=0$)  respectively. It is straightforward to check 
that the largest eigenvalue $\lambda_1 \to 0$ for larger $\varepsilon$ values in both these cases which implies that a stable/unstable stationary 
solution will retain its stability characteristics until a stability change (zero crossing of the eigenvalue/s) 
occurs in the $\varepsilon \to \infty$ limit. Furthermore using the RHC, it is easily verifiable that 
Eq.~\ref{eq:char.duffing.main2} will always have positive root/roots, whereas Eq.~\ref{eq:char.duffing.main3} will 
have all negative roots $\forall$ $\varepsilon >0$; which implies that the $x^*=0$ is always unstable and $x^*=\pm 1$ is always stable. 
Therefore, we see that partial augmentation works for stabilizing $(x^*,y^*)=(\pm 1,0)$ but \textsl{fails completely} 
to stabilize the origin $(x^*,y^*)=(0,0)$.  
Fig.~\ref{fig:figduffeigenvals} shows the largest eigenvalue calculations which verify these deductions. 
This brings us to an important observation that there might exist situations where it is not possible to target the required stationary 
solution even on using an appropriate feedback function with any combination of $k$ and $\varepsilon$ values.
\begin{figure}
\scalebox{0.495}{\includegraphics{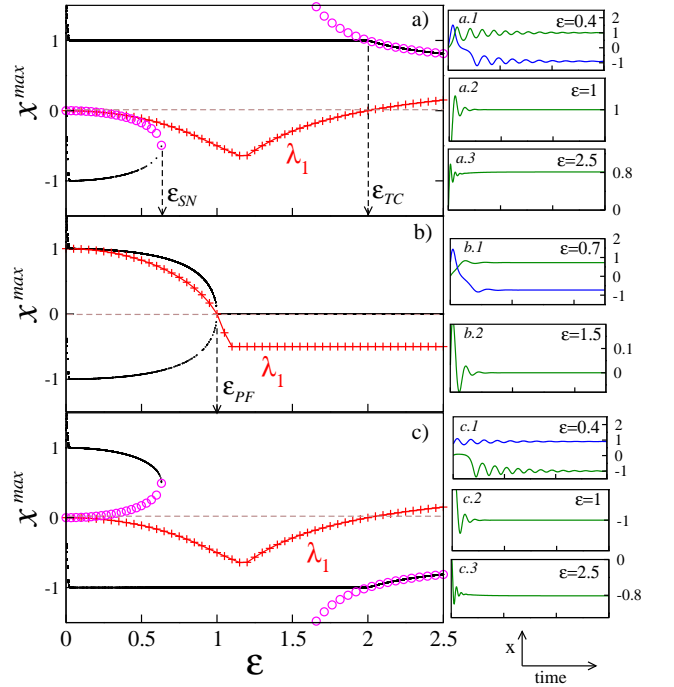}}
\caption{(Color online) Figs. a), b) and c) show the bifurcation diagrams (black 
dots) and the largest eigenvalues (red symbols) for $(x^*,y^*)=(1,0),(0,0)$, and 
$(-1,0)$ respectively. Related Figs. a.1: for $\varepsilon=0.4$, the system is 
bistable and the two related transient behaviors (in blue and green 
and likewise for other cases), a.2: for $\varepsilon=1$, the trajectory approaching 
the stable stationary solution $(1,0)$, and a.3 shows an arbitrary time series 
for $\varepsilon=2.5$. Similarly in b.1: bistability, and in b.2: the system 
approaching the stable stationary solution $(0,0)$ is shown. Identically, 
c.1, c.2, and c.3 show the bistability ($\varepsilon=0.4$), stabilization 
of $(-1,0)$ ($\varepsilon=1)$ and an arbitrary time series at $\varepsilon=2.5$.}
\label{fig:figduffing}
\end{figure}

Now considering identical augmentation with $\varepsilon_1=\varepsilon_2=\varepsilon$ and 
we will see that this system has some interesting properties.
Fig.~\ref{fig:figduffing} shows the bifurcation diagrams~\cite{bif_diagram} of the system as we 
try targeting different stationary solutions: For $(x^*,y^*)=(1,0)$, the bifurcation diagram~(black dots) is shown in 
Fig.~\ref{fig:figduffing}~a). Appropriate transient trajectories in different coupling regimes 
are also shown in related Figs.~\ref{fig:figduffing}~{(a.1), (a.2), (a.3)}. It is observed that even for 
very small coupling values, the system quickly gets into a stable stationary state regime, although, for smaller
values of $\varepsilon$, it exhibits bistability. The transient trajectories in this parameter regime are shown in 
Fig.~\ref{fig:figduffing}~(a.1). We observe that the augmentation is stabilizing our desired stationary solution 
at $(x^*,y^*)=(1,0)$, but along with it, other stationary solutions which are $\varepsilon$ dependent are also getting 
stabilized on starting with different initial conditions. 
These other stationary solutions for the augmented system here are given by,
\beqr\nonumber
{x^*}_{\pm}&=&\dfrac{1}{2}\left(-1\pm\sqrt{1-\dfrac{4\varepsilon^2}{k-\varepsilon^2}}\right),\\\nonumber
{y^*}_{\pm}&=&{x^*}_{\pm}-{{x^*}_{\pm}}^3,\\
{u^*}_{\pm}&=&-\dfrac{{y^*}_{\pm}}{\varepsilon},
\label{eq:duffing.other.solutions1}
\eqnr 
and solutions $({x^*}_{-},{y^*}_{-},{z^*}_{-})$ are observed to coexist along with $(x^*,y^*)=(1,0)$.
For higher coupling values, bistability terminates via a saddle node bifurcation when the stable branch of stationary solutions $({x^*}_{-},{y^*}_{-},{u^*}_{-})$ 
collides with the unstable branch of $({x^*}_{+},{y^*}_{+},{u^*}_{+})$(circles) as shown in in Fig.~\ref{fig:figduffing}~(a) at $\varepsilon_{SN}=\sqrt{\dfrac{k}{5}}$.
The system also exhibits hysteresis in this bistable regime and a brief discussion regarding this observation is available in Appendix~\ref{sec:duffinghysteresisappendix}. 
Beyond this regime for a range of values in $\varepsilon>\varepsilon_{SN}$, $(x^*,y^*)=(1,0)$ remains as the only stable attractor as shown in Fig.~\ref{fig:figduffing}~{(a),(a.2)}. 

In absence of augmentation, the eigenvalues for $(x^*,y^*)=(1,0)$ are complex: $\lambda_{1,2}=\pm i \sqrt{2}$. For the augmented system, 
the characteristic equation can therefore be written as:
\beqr
\lambda^3+k\lambda^2+2\lambda(1+\varepsilon^2)+(2k-\varepsilon^2)=0.
\label{eq:char.duffing1}
\eqnr
The RHC shows that this equation will have all negative roots for $(2k-\varepsilon^2)>0$ and positive root/roots appear for $\varepsilon > \sqrt{2k}$. 
This gives us the transition threshold for the destabilization of the stationary solution
as $\varepsilon^*=\sqrt{2k}$, at which the eigenvalue/s cross the zero axis. Since the 
discriminant is negative, the characteristic equation has one real and two complex conjugate roots. Considering 
large $\varepsilon$ behavior, it is seen that the largest eigenvalue $\lambda_1 \to 1/2$ which implies 
that $(x^*,y^*)=(1,0)$ is unstable in this range. The real part of the remaining complex conjugate eigenvalue pair is 
$\text{Re(}\lambda_{2,3}\text{)}= -(2k+1)/4$. At $\varepsilon^*$, Eq.~\ref{eq:char.duffing1} can be rewritten as, 
\beqr
\lambda(\lambda^2+k\lambda+2(1+2k))=0,
\label{eq:char.duffing1.1}
\eqnr
which consequently gives the eigenvalues as $\lambda_1=0$ and $\lambda_{2,3}=(-k \pm \sqrt{k^2-8(1+2k)})/2$. We can see that 
$\lambda_{2,3}$ will be a complex conjugate pair for $k \in (8-6\sqrt{2},8+6\sqrt{2})$. For our calculations, we have taken $k=2$ which 
shows that at $\varepsilon^*=\sqrt{2k}=2$, $\lambda_1$ crosses the zero line as can be seen in Fig.~\ref{fig:figduffing}~a) (red symbols). 
For higher values of $\varepsilon > \sqrt{2k}$, stationary states $({x^*}_{+},{y^*}_{+},{u^*}_{+})$ (circles) in Fig.~\ref{fig:figduffing}~(a) 
for $\varepsilon>\varepsilon^*(=\varepsilon_{TC})$ get stabilized via a transcritical bifurcation where $(x^*,y^*)= (1,0)$ and 
$({x^*}_{+},{y^*}_{+},{u^*}_{+})$ exchange their stability. This again is quite \textsl{unexpected} since the feedback is designed to 
stabilize $(x^*,y^*)= (1,0)$ for higher $\varepsilon$ values. A brief discussion regarding the behavior of this system in the 
$(\varepsilon,k)$ plane is available in Appendix~\ref{sec:duffingappendix}.

For the \textsl{origin} at $(x^*,y^*)=(0,0)$, the bifurcation diagram~(black dots) is shown in 
Fig.~\ref{fig:figduffing}~b). Appropriate transient trajectories corresponding 
to different augmentation regimes are also shown in related 
Figs.~\ref{fig:figduffing}~{(b.1), (b.2)}. We observe bistability for a range of lower
$\varepsilon$ values before the origin gets stabilized. The stationary solutions 
obtained in the bistable regime are given by
\beqr\nonumber
{x^o}_{\pm}&=&\pm \sqrt{1-\dfrac{\varepsilon^2}{k-\varepsilon^2}},\\\nonumber
{y^o}_{\pm}&=&{x^o}_{\pm}-{{x^o}_{\pm}}^3,\\
{u^o}_{\pm}&=&-\dfrac{{y^o}_{\pm}}{\varepsilon}.
\label{eq:duffing.other.solutions2}
\eqnr 
Transient trajectories in this regime demonstrating the two observed stationary solutions are 
shown in Fig.~\ref{fig:figduffing}~(b.1). These solutions approach and collapse at the origin at a pitchfork bifurcation for $\varepsilon_{PF}=\sqrt{\dfrac{k}{2}}$ 
(=1 for $k=2$ in this case) beyond which the solutions $({x^o}_{\pm},{y^o}_{\pm},{u^o}_{\pm})$ become imaginary and the origin is the only stable real stationary 
solution. A transient trajectory in this parameter regime is shown in Fig.~\ref{fig:figduffing}~(b.2).

The characteristic equation for the origin is,
\beqr
\lambda^3+k\lambda^2+\lambda(2\varepsilon^2-1)+2\varepsilon^2-k=0,
\label{eq:char.duffing2}
\eqnr
which has all negative eigenvalues for $2\varepsilon^2-k>0$ giving us a stability regime of $\varepsilon>\sqrt{k/2}$ and a transition value of 
$\varepsilon_{PF}=\varepsilon^*=\sqrt{k/2}=1$ (since $k=2$) when the eigenvalue/s cross the zero axis. 
From Eq.~\ref{eq:char.duffing2}, we get $\lambda_1=-1$ in the large $\varepsilon$ limit. 
It is numerically observed here that the discriminant $\Delta < 0$ in this range and 
therefore $\text{Re(}\lambda_{2,3}\text{)}= (1-k)/2=-0.5$ and consequently, the origin is stable in the large $\varepsilon$ limit. The largest eigenvalue 
for the origin is plotted as red symbols in Fig.~\ref{fig:figduffing}~(b) which shows the changes in the stability 
of the origin from unstable in $\varepsilon \in (0,1)$ to stable $\forall$ $\varepsilon>1$. For a discussion regarding the system 
behavior in the $(\varepsilon,k)$ plane, please see Appendix~\ref{sec:duffingappendix}.

For $(x^*,y^*)=(-1,0)$, the bifurcation diagram~(black dots) is shown in 
Fig.~\ref{fig:figduffing}~c). Appropriate transient trajectories corresponding 
to different augmentation regimes are also shown in related 
Figs.~\ref{fig:figduffing}~{(c.1), (c.2), (c.3)}. Since this solution 
is a symmetric counterpart of $(x^*,y^*)=(1,0)$, the corresponding analysis similarly carries over in this case.

These simple examples demonstrate the fact that targeting the required stationary solutions using linear augmentation 
is not quite straightforward and the procedure is quite sensitive to how the systems are augmented, the stationary solutions 
being targeted and to the properties of systems. In the following, results for a specific class of dissipative dynamical 
systems are presented to further highlight these limitations.

\subsection{Dissipative predator--prey models}\label{sec:populations}
\begin{figure*}
\scalebox{0.55}{\includegraphics{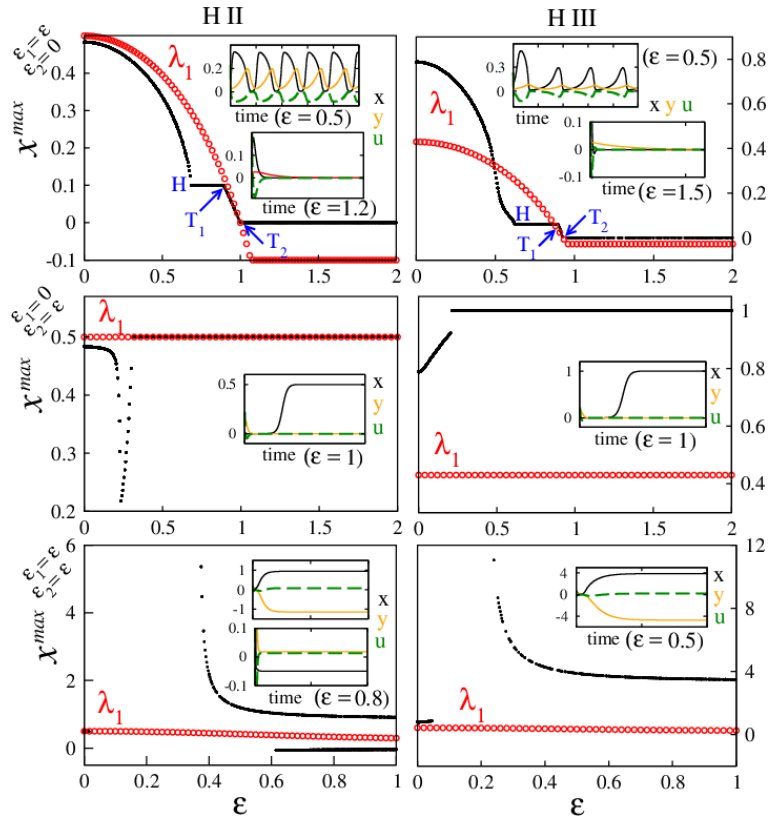}}
\caption{(Color online) Bifurcation diagram (black dots), largest eigenvalues (red circles) 
and time series (insets for specific $\varepsilon$ values) for predator--prey systems with H II (left column) 
and H III (right column) functional responses. Top row: For augmented prey, insets show the time series of 
$x,y$ and $u$ for two different $\varepsilon$ values before (with oscillatory $u$) and after the stabilization 
of $(x^*,y^*)=(0,0)$ (with $u^*=0$). The Hopf bifurcation is marked as $H$, and 
the transcritical bifurcation points have been highlighted by $T_1$, and $T_2$ respectively. Middle row: For 
augmented predator, the systems exhibit 
oscillatory behavior similar to augmented preys (not shown) for low $\varepsilon$ before they settle on the 
stationary solution $(x^*,y^*)=(K,0)$ with $u^* = 0$; where the preys reach their carrying capacity in the absence of predators 
for higher $\varepsilon$ values. Bottom row: For augmented predator and prey, for low $\varepsilon$, the systems are 
oscillatory (not shown). With increasing $\varepsilon$, both the systems 
lose the oscillatory behavior and all trajectories escape to infinity (gap in the bifurcation diagram with no black 
dots). For higher $\varepsilon$ values, unrealistic stationary solutions where either 
the preys exceed their carrying capacity $(x^*>K)$ with negative predator populations $(y^*<0)$ (H II and H III), 
or where the prey populations are negative with small positive predator population (for H II) and $u^* \ne 0$ get stabilized.}
\label{fig:populations}
\end{figure*}
Considering predator--prey population models, general evolution equations for these systems with logistic prey growth can be written as,
\beqr
\nonumber
\dot{x}& = &rx(1-x/K)-f(x)y,\\
\dot{y}& = &(\rho f(x)-\gamma)y,
\label{eq:popln1}
\eqnr
where $x$ and $y$ correspond to prey and predator populations respectively and the parameters $r, K, \rho$, and 
$\gamma$ are positive. Considering the 
evolution equation for preys, the first term $rx(1-x/K)$ represents the logistic growth 
rate of the prey species with the maximum growth rate of $r$ and carrying capacity $K$ which is the maximum population size 
that the environment can sustain indefinitely. 
The second term $f(x)y$ corresponds to the prey mortality via predation. $f(x)$ is the functional response governing 
the rate of per capita prey consumption by the predators~\cite{Solomon1949,Holling1959a,Holling1959b}. The parameter 
$\rho$ governs the 
biomass conversion efficiency for the predators in the sense of how many predators are added to the population via 
predation, and $\gamma$ is the intrinsic predator mortality parameter. 
One of the stationary solutions of this system corresponds to vanishing predator-prey populations, i.e. $(x^*,y^*) = (0,0)$. 
The other stationary solutions are dependent on the type of functional response considered. Most commonly employed $f(x)$ 
forms in such models are the Holling type with the following general expressions:
\begin{enumerate}
\item $f(x)=ax$ for Holling type I response which is identical to the predation in the Lotka--Volterra case~\cite{Lotka1910,Volterra1926},
\item $f(x)=\dfrac{ax}{(b +  x)}$ for Holling type II (Michaelis--Menten kinetics), using which, Eq.~\ref{eq:popln1} gives 
the Rosenzweig--MacArthur model~\cite{RosenzweigMacArthur1963},
\item $f(x)=\dfrac{ax^2}{(b^2 + x^2)}$ for Holling type III (Hill equation type), using which, Eq.~\ref{eq:popln1} gives the 
Truscott--Brindley model~\cite{TruscottBrindley1994} which is used in modeling phytoplankton and zooplankton interactions 
leading to harmful algal blooms,
\end{enumerate}
and consequently, corresponding stationary solutions can be obtained. The parameter $a$ 
in expressions above corresponds to the maximum per capita predation rate, and $b$ is the half 
saturation constant governing how quickly the predators attain their maximum consumption rate.
In the following, we will have a closer look at the stability properties of the trivial stationary solution 
$(x^*,y^*) = (0,0)$: \textsl{origin}. 
Considering a general augmented population model,
\beqr
\nonumber
\dot{x}& = &rx(1-x/K)-f(x)y + \varepsilon_1 u,\\\nonumber
\dot{y}& = &(\rho f(x)-\gamma)y + \varepsilon_2 u,\\
\dot{u}& = &-k u - \varepsilon_1 (x-x^*) - \varepsilon_2 (y-y^*),
\label{eq:popln2}
\eqnr
it turns out that the Jacobian for this system is identical 
for all three functional responses at the origin.
The identical characteristic equation therefore is,
\beqr\nonumber
(r-\lambda)(\gamma+\lambda)(k+\lambda)+{\varepsilon_2}^2(r-\lambda)-{\varepsilon_1}^2(\gamma+\lambda)=0.\\
\label{eq:char.popln}
\eqnr
For $\varepsilon_1=\varepsilon_2=0$, we obtain the eigenvalues as $\lambda_1=r, \lambda_2=-\gamma$, 
and $\lambda_3=-k$ where $\lambda_1$ and $\lambda_2$ are the eigenvalues for the original system in 
Eq.~\ref{eq:popln1} implying that the origin is unstable, and $\lambda_3$ corresponds to the exponentially 
decaying augmentation variable $u$. Since the Holling type I case with insatiable predators is quite unrealistic, 
we will focus here on systems with Holling type II (H II) and III (H III) behaviors. In the 
following analysis, the parameter values are fixed at: $r = 0.5$, $K = 0.5$, $a = 1/3$, $b = 1/15$, 
$\rho = 0.5$, $\gamma = 0.1$ for the H II~\cite{Roos2013} system, and $r = 0.43$, $K = 1$, $a = 1$, $b= 0.053$, $\rho = 0.05$, 
$\gamma = 0.028$ for H III~\cite{TruscottBrindley1994,Karnatak2014a}. Let us now look at different augmentation situations:

For $\varepsilon_1=\varepsilon$ and $\varepsilon_2=0$, i.e. only prey populations are augmented, substituting these values 
in Eq.~\ref{eq:char.popln} gives, 
\beqr
(\gamma+\lambda)[(r-\lambda)(k+\lambda)-{\varepsilon}^2]=0.
\label{eq:char.popln1}
\eqnr
Since one of the roots $\lambda=-\gamma$ is independent of $\varepsilon$ therefore the remaining roots of this equation 
determine the stability of the origin. 
The remaining two roots are $\lambda_{\pm}= -(k-r)/2\pm\sqrt{(k+r)^2-4\varepsilon^2}/2$ out of which, $\lambda_- <0$, 
$\forall$ $\varepsilon$. It is easily verifiable 
that the eigenvalue $\lambda_+$ (which also is the largest) is positive $\forall$ $\varepsilon < \sqrt{kr}$ and crosses the zero 
axis at $\varepsilon^*=\sqrt{kr}$ leading to all negative eigenvalues and hence a stable origin. 
This is quite similar to the harmonic oscillator case where increasing/decreasing the value of the decay parameter 
$k$ could increase/decrease the threshold value of stable $\to$ unstable transition (unstable $\to$ stable in this case). 
Furthermore, in the large $\varepsilon$ limit, 
we obtain the largest eigenvalue $\lambda_1 = -\gamma$ and therefore the origin is stable in this regime. 
Fig.~\ref{fig:populations}: top row shows the bifurcation diagram~\cite{bif_diagram} and the largest eigenvalue 
behavior for H~II (left) and III (right). The unstable $\to$ stable 
transition in both these systems with increasing coupling values can be seen in the figure. Although for the 
H III system in the regime $\varepsilon > \sqrt{rk} (= 0.927,$ for $r=0.43, k=2)$, certain initial conditions lead to the 
trajectories escaping to infinity (not shown) which accounts for the
missing dots in the bifurcation figure. Since $x$ and $y$ are population variables by definition, population models are 
constrained to work for non-negative values of $x$ and $y$ respectively. What we observe here is that the augmentation forces the prey 
populations into taking negative values which leads to a breakdown in the model constraints and the logistic function in the rate equation of $x$ 
leads to diverging solutions as time increases.  For H II system this appears not to be the case and all considered initial conditions 
lead to a stable origin $\forall$ $\varepsilon > \sqrt{rk} (= 1,$ for $r=0.5, k=2)$.

For $\varepsilon_1=0$ and $\varepsilon_2=\varepsilon$, i.e. only the predator populations are augmented, substituting these 
values in Eq.~\ref{eq:char.popln} give,
\beqr
(r-\lambda)[(\gamma+\lambda)(k+\lambda)+{\varepsilon}^2]=0,
\label{eq:char.popln2}
\eqnr
and we see that an eigenvalue $\lambda=r$ is always positive since $r>0$, and therefore this setup will \textsl{never} stabilize 
the origin. The remaining eigenvalues are 
$\lambda_{\pm}=-(\gamma+k)/2 \pm \sqrt{(k-\gamma)+4\varepsilon^2}/2$. In Fig.~\ref{fig:populations}: second row, for low $\varepsilon$ 
values, the systems exhibit periodic 
behavior similar to the one shown for the augmented prey case. For higher values of $\varepsilon$, both H~II and H~III settle on a 
stationary solution of the original system $(x^*, y^*) = (K,0)$ which we did not intent to stabilize. For this solution, the preys 
exist at their carrying capacity and the predators vanish. For H II and H III, the carrying 
capacities considered for simulations are $K=0.5$ and $K=1$ respectively, and hence the observations in Fig.~\ref{fig:populations} (middle row).

For $\varepsilon_1=\varepsilon_2=\varepsilon$, i.e. both prey and predator populations are augmented, substituting these values in 
Eq.~\ref{eq:char.popln} and rearranging terms gives,
\beqr\nonumber
\lambda^3+(\gamma+k-r)\lambda^2+(2\varepsilon^2+(\gamma-r)k-r\gamma)\lambda\\
+\varepsilon^2(\gamma-r)-r\gamma k=0.
\label{eq:char.popln3}
\eqnr
Using the RHC~\cite{Routh_Hurwitz_criteria}, one of the conditions for this equation to have all negative roots 
is $\varepsilon>\sqrt{\dfrac{r\gamma k}{(\gamma - r)}}$ which is impossible to achieve 
since $r>\gamma$. Therefore, this setup will not stabilize the origin either. Fig.~\ref{fig:populations}: 
bottom row shows the behavior of H II and H III. For smaller $\varepsilon$ values, these systems 
exhibit periodic behavior similar to the augmented prey. On increasing $\varepsilon$ further, systems 
enter a regime where all considered initial conditions lead to escaping trajectories. The reason behind this behavior here again is due 
to a breakdown in modeling constraints. 
Examination of transient trajectories reveals that augmentation in this case is forcing the predator populations into $y<0$ axis which leads to a breakdown in the model, thereby 
initiating a positive feedback loop in the prey populations leading to the diverging behaviors observed in simulations. 
Beyond this regime for higher values of $\varepsilon$, H II system exhibits 
bistability between different stationary solutions where in one case, preys exceed their carrying capacity $(x^*> K)$ and 
the predator populations are negative $(y^*<0)$, and in the other case, the prey populations are negative $(x^*< 0)$ and 
predators assume a small positive value. It is important to note yet again that these solutions are impractical because the populations 
cannot exist above their carrying capacities nor can they take negative values under realistic modeling constraints. 
For H III system, we only observe the equilibrium 
solutions with $x^*> K$ and $y^*<0$ (see inset). In both the cases, we have a non vanishing $u^*>0$ and therefore these solutions exist 
due to augmentation and cannot be observed otherwise.
Following this analysis, we can conclude that augmenting the prey is the correct strategy to stabilize of the origin and 
the other coupling schemes can lead to complicated dynamics. Even though the analysis here is limited to the 
origin, we can expect these behaviors to be quite general with respect to other stationary solutions as well. 
 \begin{figure}
\scalebox{0.28}{\includegraphics{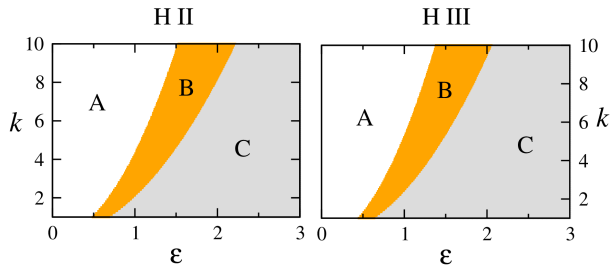}}
\caption{(Color online) Different dynamical regimes for H II and H III systems are marked as \textsl{A, B} and \textsl{C} 
in the $\varepsilon$, $k$ plane. \textsl{A} is
the regime of periodic dynamics, in \textsl{B} stationary solutions of the coupled system are stable and \textsl{C} is the regime 
of stable origin. The boundaries between A$\to$B and B$\to$C are the loci of the reverse Hopf 
bifurcation $H$ and the second transcritical bifurcation $T_2$ respectively (as in Fig.~\ref{fig:populations}: top row).}
\label{fig:parscans}
\end{figure}

Now considering applications, as already mentioned, origin corresponds to an equilibrium for which the predators 
and the preys vanish. Persistence of populations for a proper ecosystem function is very imperative and has been studied 
extensively from several perspectives, contributing towards a better understanding of the processes leading to 
species extinction~\cite{Lande1993,Hanski1999,Hughes1997,Ceballos2002,Folke2004}. Knowledge regarding these processes can 
help in devising procedures which can contribute towards better species conservation efforts. For the simple models 
considered in the previous analysis, it is clear that either by coupling the system appropriately or by using specific parameter values for $k$ and $\varepsilon$, 
we can avoid stabilizing the origin. For instance, considering the prey augmented case, for low $\varepsilon$, the systems exhibit periodic oscillations. 
On increasing the coupling strength, stationary solutions of the augmented system $(x^*>0,y^*>0,u^*<0)$, satisfying
\beqr
\nonumber
rx^*(1-x^*/K)-f(x^*)y^* + \varepsilon_1 u^* &=& 0,\\\nonumber
(\rho f(x^*)-\gamma)y^* &=& 0,\\
-k u^* - \varepsilon x^* &=& 0,
\eqnr
get stabilized through a reverse Hopf bifurcation (marked as $H$ in Figs.~\ref{fig:populations} (top row)). For these 
stationary solutions, the value of $x^*$ stays constant while $y^*$ and $u^*=-\varepsilon x^*/k$ show a variation for a range 
of $\varepsilon$ values (plateau between $H$ and $T_1$ in Fig.~\ref{fig:populations} (top row)). 
It is also important to note that some initial conditions in this regime can lead to trajectories escaping to infinity. This branch 
of solutions undergoes a transcritical bifurcation (marked $T_1$ in Fig.~\ref{fig:populations} (top row)) where 
it exchanges stability with another branch of solutions with $u^* \to 0$ for increasing $\varepsilon$. 
At $\varepsilon^*=\sqrt{rk}$, $u^*=0$ and the predator--prey system effectively decouples from augmentation which is accompanied by another 
transcritical bifurcation ($T_2$ in Fig.~\ref{fig:populations} (top row)) between 
the continuing branch of stationary solutions $(x^*>0,y^*>0,u^*\to 0)$ and the origin. In $\varepsilon>\varepsilon^*$ 
regime, origin is the only dynamical attractor. Fig.~\ref{fig:parscans} shows the parameter scans for H II and H III 
systems highlighting these different dynamical regimes. In region \textsl{A} these systems exhibit periodic behavior and 
the boundary between \textsl{A} and \textsl{B} is the locus of the Hopf bifurcation in the $\varepsilon$, $k$ plane, 
which leads to the stabilization of stationary solutions $(x^*>0,y^*>0,u^*<0)$. \textsl{B} corresponds to the regime where 
stable stationary solutions $(x^*>0,y^*>0,u^*<0)$ and $(x^*>0,y^*>0,u^*\to 0)$ are observed and the boundary between 
\textsl{B} and \textsl{C} is the locus of the second transcritical bifurcation $T_2$ which leads to the stabilization of the origin. 
Therefore by using appropriate values of $\varepsilon$ and $k$, we can keep the system in either a periodic 
state, or a stationary state with non vanishing populations and can expect this procedure to work in 
experiments and be robust with respect to noise; Ref.~\cite{Pooja2011} experimentally stabilized 
a stationary solution in an electronic Lorenz system at permitted noise level. Furthermore, in the other instances of augmented 
predators, or augmented predators and preys we already observe a complete lack of origin stabilization. Therefore, we can employ these 
schemes as well to avoid stabilizing the origin but one needs to be careful since these cases can lead to other complications as discussed. 
Another useful application for these observations could be in cases where maximization of prey yield is required. Augmenting 
 the predator populations is seen to stabilize the equilibrium where the prey populations exist at their carrying capacity and 
 the predators vanish. This can find applications in fisheries~\cite{Yodzis1994,Christensen1996}, algae fuel 
 generation~\cite{Scott2010,Oncel2013}; where maximal sustainable yields are crucial, and also in biomedical research, for e.g. in HIV-1 infection 
 models~\cite{Boer1998} where a portion of human immune system i.e. activated CD$4^{+}$ T cells are the primary target of the HIV-1 infection~\cite{Fauci1993,Weiss1993} 
 which can be modeled via predator--prey dynamics.

\section{Summary}\label{sec:summary}
In this work we studied the general ability of linear augmentation towards stabilizing desired stationary solutions of oscillatory systems. Through some
simple examples discussed in this paper, it is clear that the effectiveness of this scheme 
is quite sensitive to the augmentation parameters, the class of oscillatory systems considered, the stationary solutions to stabilize and also on the way 
the systems are augmented. 
Therefore, although the simplicity of linear augmentation makes it a very compelling choice for applications, a careful analysis is required 
to test the system for potential pitfalls associated with the scheme. As highlighted by the examples, apart from failing to
target the appropriate stationary solutions, linear augmentation can also lead to other complicated dynamical situations which include escaping 
trajectories, stabilization of unintended stationary solutions or the stabilization of stationary solutions which are not permitted under the 
modeling constraints; preys existing above their carrying capacities and negative predator populations in Sec.~\ref{sec:populations} for instance.
Nevertheless, one can find ways to exploit the failures of the scheme in applications. Although we can 
expect to see these results in experiments, an in-depth study of this procedure in presence of noise, and also for larger systems
is required. Extending on the results in the ecological context, one needs to check the process behavior in presence of multiple preys and predators, for a food chain, 
and also for other functional responses~\cite{Karnatak2015}. Furthermore, linear augmentation has been proposed as a mechanism to 
control bistability~\cite{Pooja2013} but how it fares in managing more general instances of 
multistability including extreme multistability~\cite{Ngonghala2011, Karnatak2014a} is still an open question and will be addressed 
in subsequent studies~\cite{Karnatak2015}.\\

{\bf Acknowledgment:} The author would like to thank anonymous referees for their constructive comments and suggestions, A. Nandi, S. Bialonski for a critical reading 
of the manuscript and related discussions, and A. V. Costea for help with the text. The author acknowledges the Max Planck Society for research support.

\appendix
\section{Converging and diverging trajectories in augmented harmonic oscillator}\label{sec:harmonicappendix}
\begin{figure}
\scalebox{0.5}{\includegraphics{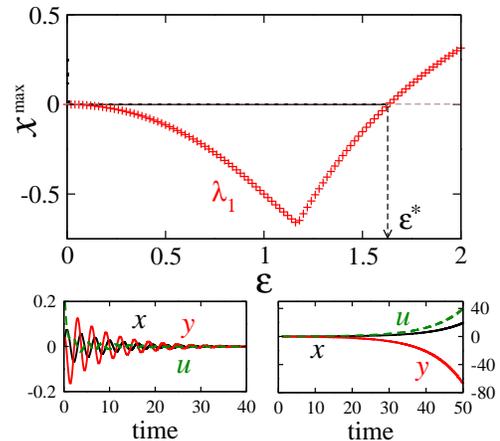}}
\caption{(Color online) Bifurcation diagram (black dots) along with the largest eigenvalue (red symbols) as a function of $\varepsilon$ in the top row. 
$\varepsilon^*=\omega\sqrt{\dfrac{k}{\omega^2-1}}(\sim 1.63)$ marks the coupling beyond which all initial 
conditions lead to escaping trajectories. Transient trajectories shown for $\varepsilon(=0.5)<\varepsilon^*$ (left bottom) and 
$\varepsilon(=1.75)>\varepsilon^*$ (right bottom).}
\label{fig:harmonicappendix1}
\end{figure}
Augmented harmonic oscillator dynamics from Eq.~\ref{eq:harmonic2} can also be expressed in form of a second order ODE as,
\beqr
\mathbb{D}x=U(\varepsilon_1,\varepsilon_2,k, t)[=\lbrace\varepsilon_2+\varepsilon_2{\varepsilon_1}^2-\varepsilon_1 k\rbrace u(t)],
\label{eq:2orderfulharmonic1}
\eqnr
where the derivative operator $\mathbb{D}=D^2+\varepsilon_1\varepsilon_2D+({\varepsilon_1^2}+\omega^2)$ with $D^i=\dfrac{d^i}{dt^i}$, $i=1,2$ 
in this case. This equation corresponds to a driven harmonic oscillator with frequency $(\varepsilon_1^2+\omega^2)$ and a damping coefficient 
$\varepsilon_1\varepsilon_2$. The roots of the auxiliary equation for the operator $\mathbb{D}$ are $m_{\pm}=\alpha\pm\beta$ 
where $\alpha=-{\varepsilon_1\varepsilon_2}/2$ and 
$\beta=\sqrt{\varepsilon_1^2 \varepsilon_2^2-4(\varepsilon_1^2+\omega^2)}/2$. For partially augmented cases 
$\alpha=0$ and $\beta=\sqrt{-4(\varepsilon_1^2+\omega^2)}/2$ or $\beta=i\omega$ for $\varepsilon_2=0,\varepsilon_1 \ne 0$ 
and $\varepsilon_1=0, \varepsilon_2 \ne 0$ respectively.

For identical augmentation $\varepsilon_1=\varepsilon_2=\varepsilon$, we get $\mathbb{D}=D^2+\varepsilon^2D+(\varepsilon^2+\omega^2)$ and Eq.~\ref{eq:2orderfulharmonic1} reads
\beqr
\mathbb{D}x=U(\varepsilon,k,t)[=\varepsilon\lbrace1+\varepsilon^2-k\rbrace u(t)].
\label{eq:2orderfulharmonic2}
\eqnr
The roots of the auxilliary equation in this case are $m_{\pm}=\alpha\pm\beta$ with $\alpha=-{\varepsilon^2}/2$ and 
$\beta=\sqrt{\varepsilon^4-4(\varepsilon^2+\omega^2)}/2$.
For imaginary $\beta$, the transient solution for Eq.~\ref{eq:2orderfulharmonic2} can be expressed as,
\beqr
x_g(t)=A x_1(t)+B x_2(t),
\label{eq:sol1fulharmonic}
\eqnr
which is independent of the forcing term $U(\varepsilon,k,t)$ with $x_1(t)=\exp{(\alpha t)}\cos{\beta t}$, and $x_2(t)=\exp{(\alpha t)}\sin{\beta t}$. 
Consequently, the steady state solution can be obtained by using the Laplace and inverse Laplace transformations giving,
\beqr\nonumber
x_{st}(t)=\dfrac{1}{\Omega}\int_{0}^{t}e^{-\gamma(t-x)}\sin{(\Omega(t-x))} U(\varepsilon, k,x)dx,\\
\label{eq:sol2fulharmonic}
\eqnr
where $\Omega=\sqrt{\omega_0^2-\gamma^2}$, ${\omega_0}^2=\varepsilon^2+\omega^2$ and $\gamma=\varepsilon^2/2$.
Now at this point, we do not know the exact expression for $U(\varepsilon,k,t)$. 
Considering the transient behavior of trajectories in partially/fully 
augmented system, we clearly observe that they possess an exponentially decaying/diverging envelop (see Fig.~\ref{fig:figharmonic}~(inset) 
and Fig.~\ref{fig:harmonicappendix1} bottom row). Based on these observations, assuming $U(\varepsilon,k,t)=a_0 \exp{(k_m t)}$ where 
both $a_0,k_m$ are functions of $\varepsilon$ and $k$, and solving Eq.~\ref{eq:sol2fulharmonic} gives
\beqr
x_p(t)=\exp{(k_mt)}\left(\dfrac{a_0}{(k_m+\alpha)^2-\beta^2}\right).
\label{eq:sol3fulharmonic}
\eqnr
From this expression we see that the trajectories will exponentially decay to the origin $\forall$ $k_m<0$ and diverge $\forall$ $k_m>0$.
\begin{figure}
\scalebox{0.4}{\includegraphics{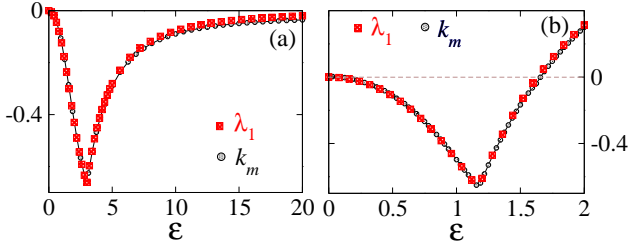}}
\caption{(Color online) Largest eigenvalues and estimated values of $k_m$ as functions of $\varepsilon$, (a) for the system augmented in $x$ and (b) for the fully augmented system.}
\label{fig:harmonicappendix2}
\end{figure}

Fig.~\ref{fig:harmonicappendix2} shows the numerical estimation of $k_m$ along with the largest eigenvalue of the Jacobian $\lambda_1$ at the origin as a function of $\varepsilon$; 
for partially (Fig.~\ref{fig:harmonicappendix2}~(a)) and fully augmented cases (Fig.~\ref{fig:harmonicappendix2}~(b)). $k_m$ here was calculated as the average rate of 
convergence/divergence in the Euclidean distance of the current systems' state from its previous state, for every time step along the trajectory. These results suggest that
$k_m = \lambda_1$ and this observation has some interesting consequences. For $k_m= \lambda_1<0$, we have a case of a harmonic oscillator being driven by an exponentially decaying 
force and consequently the system settles on the origin as $t \to \infty$. Similarly the other case of an unstable origin ($k_m= \lambda_1>0$) is equivalent to the oscillator under
the influence of an exponentially diverging force (energy being pumped into the system) which leads to diverging trajectories as time increases.
\section{Hysteresis in augmented Duffing model}\label{sec:duffinghysteresisappendix}
\begin{figure}
\scalebox{0.48}{\includegraphics{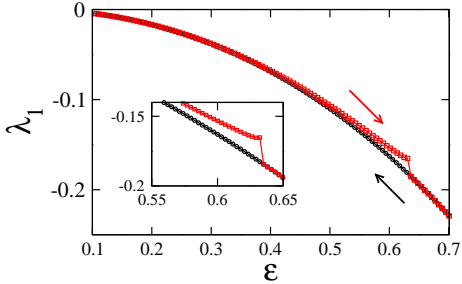}}
\caption{(Color online) Maximal Lyapunov exponent for increasing and decreasing values of $\varepsilon$ for the fully augmented Duffing system with $(x^*=1,y^*=0)$
demonstrating hysteresis. Calculations for increasing and decreasing $\varepsilon$ are marked by red and black arrows respectively.}
\label{fig:hysteresis-duffing}
\end{figure}
For the bistable fully identically augmented Duffing system in Eq.~\ref{eq:duffing1} with $x^*=1,y^*=0$, the largest Lyapunov exponent was calculated for increasing 
and decreasing values of augmentation strength $\varepsilon$. Starting with initial conditions leading to the solutions $({x^*}_{-},{y^*}_{-},{z^*}_{-})$ at $\varepsilon=0.1$, the 
initial conditions for the next calculation at $\varepsilon=0.1+\delta\varepsilon$ were taken as the final values of $x,y,u$ from the previous calculation for $\varepsilon=0.1$, with 
$\delta\varepsilon=0.001$ and so on for the entire range in the forward direction. Similarly for backwards calculation, the process was repeated starting from $\varepsilon=0.7$ where $(1,0)$ 
is the only stable solution with $\delta\varepsilon=-0.001$. The results of the calculation are shown in Fig.~\ref{fig:hysteresis-duffing} and as one would expect, this system exhibits 
hysteresis in the interval of bistablity. 
\section{Fully augmented Duffing system: behavior in $(\varepsilon,k)$ plane}\label{sec:duffingappendix}
\begin{figure}
\scalebox{0.33}{\includegraphics{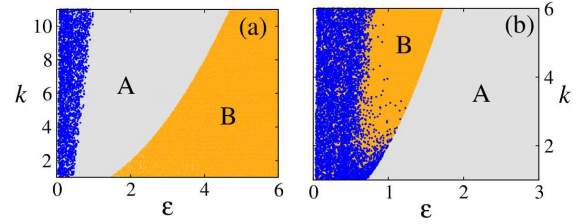}}
\caption{(Color online) Behavior of fully augmented Duffing system in the $(\varepsilon,k)$ plane. Parameter 
regimes marked in grey (\textbf{A}) correspond to a successful stabilization of the intended stationary solution; $(1,0)$ in (a) and $(0,0)$ in (b). 
Regimes of bistability are marked with blue dots and \textbf{B} corresponds to parameter values for which other stationary solutions are stable.}
\label{fig:parscans-duffing}
\end{figure}
Different dynamical regimes
for the Duffing system in Eq.~\ref{eq:duffing1} with $(x^*,y^*)=(1,0)$ and $(0,0)$ are shown in Fig.~\ref{fig:parscans-duffing}. The grey areas marked as \textbf{A} correspond 
to the regimes where the intended stationary solutions are successfully stabilized, namely $(1,0)$ and $(0,0)$ in Figs.~\ref{fig:parscans-duffing}(a) 
and (b) respectively. 

For $x^*=1,y^*=0$ in Fig.~\ref{fig:parscans-duffing}(a): blue dots for lower $\varepsilon$ values highlight the regimes of bistability where solutions 
$({x^*}_{-},{y^*}_{-},{z^*}_{-})$ (from Eq.~\ref{eq:duffing.other.solutions1}) and $(1,0)$ coexist. Solutions $({x^*}_{-},{y^*}_{-},{z^*}_{-})$ vanish via 
a saddle node bifurcation after colliding with the unstable branch of solutions $({x^*}_{+},{y^*}_{+},{z^*}_{+})$ (again from Eq.~\ref{eq:duffing.other.solutions1}) 
and the boundary of the blue dot regime gives the locus of this saddle node bifurcation; parabolic function $F_{SN}(\varepsilon,k)(=5\varepsilon^2-k)=0$, estimated 
from the expression for ${x^*}_{\pm}$ from Eq.~\ref{eq:duffing.other.solutions1} as the limiting value of $k$ and $\varepsilon$ to get real solutions, obtained by 
equating the discriminant in the expression of ${x^*}_{\pm}$ to zero.
The boundary between regimes \textbf{A} and \textbf{B} corresponds to the locus of the transcritical bifurcation between $(1,0)$ and 
$({x^*}_{+},{y^*}_{+},{z^*}_{+})$ given by the zero crossing of the largest eigenvalue for $(1,0)$ which satisfies $F_{TC}(\varepsilon,k)(=2\varepsilon^2-k)=0$.
In region \textbf{B} either the now stable coupling dependent stationary solutions
$({x^*}_{+},{y^*}_{+},{z^*}_{+})$, or escaping trajectories are observed.

Similarly for $x^*=0,y^*=0$ in Fig.~\ref{fig:parscans-duffing}(b): \textbf{B} highlights the regime where the system settles on the solutions $({x^o}_{+},{y^o}_{+},{u^o}_{+})$. 
The blue dots correspond to the initial conditions leading to stationary solutions $({x^o}_{-},{y^o}_{-},{u^o}_{-})$. Since the system is bistable in this regime, 
we should expect the entire region \textbf{B} to be densely filled with these blue dots but that is not the case. The reason behind this behavior is a difference in the relative basin size of 
these two solutions; number of initial conditions leading to $({x^o}_{+},{y^o}_{+},{u^o}_{+})$ is more than the ones which lead 
to $({x^o}_{-},{y^o}_{-},{u^o}_{-})$. This difference is even more pronounced for higher values of $k$. Both these solutions vanish via a pitchfork bifurcation and the locus of this 
bifurcation which separates regimes \textbf{B} and \textbf{A} can be traced by the function 
$F_{PF}(\varepsilon,k)(=2\varepsilon^2-k)=0$ which is obtained by equating the discriminant in the expression of ${x^o}_{\pm}$ from Eq.~\ref{eq:duffing.other.solutions2} 
to zero.


\begin{thebibliography}{56}
\bibitem [{bif()}]{bif_diagram}%
  \BibitemOpen
  \href@noop {} {}\bibinfo {note} {The system is evolved for 10 different
  initial condition for each value of the control parameter sufficiently for
  the transients to be discarded. System is then further evolved to capture the
  dynamical regimes in the system by noting all possible extrema in the
  dynamics for the duration of evolution}\BibitemShut {NoStop}%
\bibitem [{Rou()}]{Routh_Hurwitz_criteria}%
  \BibitemOpen
  \href@noop {} {}\bibinfo {note} {Given a polynomial $P(\lambda)=\lambda^n +
  a_1 \lambda^{n-1} + \ldots +a _{n-1}\lambda + a_n$, where the coefficients
  $a_i$'s are real, then all roots of the polynomial are negative or have
  negative real parts iff the determinants of all Hurwitz matrices are
  positive. For $n=3$, this occurs when the coefficients $a_1, a_2$ and $a_3$
  are strictly positive and $a_1a_2>a_3$.}\BibitemShut {Stop}%
\bibitem [{dis()}]{discriminant}%
  \BibitemOpen
  \href@noop {} {}\bibinfo {note} {For any general cubic equation
  $f(x)=ax^3+bx^2+cx+d$, the discriminant is defined as
  $\Delta=18abcd-4b^3d+b^2c^2-4ac^3-27a^2d^2$.}\BibitemShut {Stop}%
\end{thebibliography}
\end{document}